\documentclass[conference]{IEEEtran}

\usepackage[textwidth=1cm,backgroundcolor=yellow,textsize=small]{todonotes}

\usepackage{url}
\usepackage[noadjust]{cite}
\usepackage{multirow} 
\usepackage{algpseudocode}
\usepackage{algorithm}
\usepackage{rotating}
\usepackage{kantlipsum} 
\usepackage{commath}
\allowdisplaybreaks
\usepackage{mathtools}  
\usepackage{verbatim}
\usepackage{xr-hyper} 
\usepackage{enumitem}

\usepackage{epstopdf}
\usepackage{wrapfig}
\usepackage{latexsym}
\usepackage{amssymb}
\usepackage{amsthm}
\usepackage{amsfonts}
\usepackage{amsmath} 
\usepackage{graphicx}
\usepackage{latexsym}
\usepackage{booktabs}
\usepackage{support-caption} 
\usepackage{subcaption} 
\usepackage{xspace}

\usepackage[normalem]{ulem} 

\usepackage[T1]{fontenc}

\ifCLASSINFOpdf
\else
\fi
\usepackage{tikz}
\usetikzlibrary{positioning, shapes, arrows}

\newcommand{\dq}[1]{``#1''}

\newif\ifcommentson
\commentsontrue

\newif\ifextended
\newif\ifshortver

\extendedtrue

\newcommand{\extended}[1]{\ifextended \ifshortver \textcolor{purple}{#1} \else \textcolor{black}{#1} \fi  \fi}

\newif\ifrevision

\begin{document}

\bstctlcite{IEEEexample:BSTcontrol}



\title{Sharing GPUs and Programmable Switches in a Federated Testbed with SHARY}

\author{\IEEEauthorblockN{
Stefano Salsano\IEEEauthorrefmark{1}\IEEEauthorrefmark{2},
Andrea Mayer\IEEEauthorrefmark{1}\IEEEauthorrefmark{2},
Paolo Lungaroni\IEEEauthorrefmark{1},\\
Pierpaolo Loreti\IEEEauthorrefmark{1}\IEEEauthorrefmark{2},
Lorenzo Bracciale\IEEEauthorrefmark{1}\IEEEauthorrefmark{2},
Andrea Detti\IEEEauthorrefmark{1}\IEEEauthorrefmark{2},
Marco Orazi\IEEEauthorrefmark{1},\\
Paolo Giaccone\IEEEauthorrefmark{3},
Fulvio Risso\IEEEauthorrefmark{3},
Alessandro Cornacchia\IEEEauthorrefmark{4},
Carla Fabiana Chiasserini\IEEEauthorrefmark{3}
}
\IEEEauthorblockA{
\IEEEauthorrefmark{1}University of Rome Tor Vergata,
\IEEEauthorrefmark{2}CNIT,
\IEEEauthorrefmark{3}Politecnico di Torino,\\
\IEEEauthorrefmark{4}KAUST (King Abdullah University of Science and Technology)
}
\vspace{2ex}
\vspace{-4ex}
\\ 
\extended{\textbf{Extended version of the paper accepted to NOMS 2025 - v01 - January 2025}}
}

\markboth{Journal of \LaTeX\ Class Files,~Vol.~14, No.~8, August~2015}%
{Shell \MakeLowercase{\textit{et al.}}: Bare Demo of IEEEtran.cls for IEEE Journals}
%



\maketitle

\begin{abstract}
Federated testbeds enable collaborative research by providing access to diverse resources, including computing power, storage, and specialized hardware like GPUs, programmable switches and smart Network Interface Cards (NICs). Efficiently sharing these resources across federated institutions is challenging, particularly when resources are scarce and costly. GPUs are crucial for AI and machine learning research, but their high demand and expense make efficient management essential. Similarly, advanced experimentation on programmable data plane requires very expensive programmable switches (e.g., based on P4) and smart NICs. 

This paper introduces SHARY (SHaring Any Resource made easY), a dynamic reservation system that simplifies resource booking and management in federated environments. We show that SHARY can be adopted for heterogenous resources, thanks to an adaptation layer tailored for the specific resource considered. Indeed, it can be integrated with FIGO (Federated Infrastructure for GPU Orchestration), which enhances GPU availability through a demand-driven sharing model. By enabling real-time resource sharing and a flexible booking system, FIGO improves access to GPUs, reduces costs, and accelerates research progress. SHARY can be also integrated with SUP4RNET platform to reserve the access of P4 switches.
\end{abstract}


%
\IEEEpeerreviewmaketitle

\section{Introduction}
%
%
%
%


Federated testbeds have emerged as powerful platforms for supporting research and development in a variety of fields, including cloud computing, networking, and artificial intelligence. These testbeds allow researchers to access and use a diverse array of resources — such as compute power, storage, and networking devices — at geographically distributed sites. However, despite the advantages that federated testbeds provide, managing and sharing resources effectively remains a significant challenge. The complexity of integrating heterogeneous resources like GPUs, physical networking devices such as SmartNICs or programmable switches (e.g., P4-based switches), and general-purpose compute nodes often leads to inefficiencies in utilization and availability.

In this work, we focus on a federated testbed within the context of the Italian RESTART research program, “RESearch and innovation on future Telecommunications systems and networks, to make Italy more smart”~\cite{restartWebsite}. This program is the most significant public \textit{R\&D} initiative ever implemented in the Telecommunications sector in Italy. In the context of the RESTART program, our goal is to develop tools and frameworks that create a more adaptive and efficient resource-sharing ecosystem, better aligned with the needs of modern research workflows, and to support the scientific community in leveraging the full potential of the available infrastructure.

\subsection{Sharing resources in federated testbeds}
One of the core issues in federated environments is the \textit{under-utilization of specialized hardware resources}, such as GPUs or programmable switches. While GPUs are crucial for accelerating machine learning and artificial intelligence workloads, they are often idle when allocated exclusively to a single researcher or project. The same applies to physical networking devices like programmable switches and SmartNICs, which are frequently underused due to their specificity and the lack of tools that enable dynamic sharing among multiple users or projects. These inefficiencies are further compounded by the high acquisition costs associated with all these resources. The importance of achieving high utilization is particularly critical for smaller-sized federated testbeds and, more generally, when the number of available resources is limited either in absolute terms or relative to the demand.

Coordinating access to shared resources in a fair and efficient manner is essential to achieving these goals. Traditional resource management approaches, such as static partitions or fixed time slots, often fail to adapt to the dynamic nature of research workloads, leading to resource contention or wasted capacity. Federated testbeds need mechanisms that can dynamically allocate resources based on real-time demand while ensuring that the reservation process remains transparent and user-friendly. In practical terms, this means that the basic time-slot-based reservation mechanisms need to be complemented by dynamic re-allocation procedures based on actual usage and demand. This approach ensures that critical resources (such as programmable switches, SmartNICs, GPUs) are fully leveraged, maximizing the value of the investment and providing broader access to researchers. 

Another major challenge in federated testbeds is the \textit{heterogeneity of resources and interfaces}. Each site within a federated network may have different types of hardware, ranging from traditional CPUs to high-performance GPUs, specialized FPGA accelerators, and various networking devices. This diversity in hardware introduces complexities in the management of resources, as different devices often require different drivers, software environments, and management protocols. 


A further issue to be addressed is the ability to support experiments that span multiple types of resources, such as combining GPUs for computation with specialized networking devices for data transfer. These types of experiments often require precise synchronization and integration of resources from multiple sites, making the coordination of heterogeneous resources even more challenging. The complexity of orchestrating different types of hardware and ensuring that they work seamlessly together is crucial to enabling advanced research scenarios, where diverse computational and network requirements must be met simultaneously.

\subsection{Sharing GPUs in federated testbeds}

Focusing on Graphics Processing Units (GPUs), they have become indispensable assets in the realm of artificial intelligence (AI) and machine learning research. Their parallel processing capabilities significantly accelerate the training and inference of complex models. However, the burgeoning demand for GPUs has outpaced their supply, resulting in a notable scarcity and elevated costs. This situation poses a considerable challenge, particularly for academic and research institutions with limited budgets. Researchers typically seek to access the latest GPU architectures to push AI performance to their limit or faithfully reproduce production training environments. In contrast, new GPU models are provisioned gradually and usually limited in quantity, generating substantial demand for a largely oversubscribed pool of resources. This represents a critical bottleneck, unlike in CPU sharing environments where users typically show less concern about the specific CPU model utilized.

In many scenarios, individual researchers or small teams acquire GPUs for their exclusive use. Despite the high demand, these GPUs often remain underutilized due to the nature of research workflows, where periods of intensive computation are interspersed with idle times. The implementation of a federated system in which GPU resources can be shared among researchers can mitigate this problem. Such a system would enable for more efficient utilization of available GPUs, distributing computational loads dynamically based on real-time demand. 

Current open-source virtualization solutions for GPU resource allocation rely predominantly on static partitioning methods. These methods allocate fixed portions of GPU resources to users or tasks, regardless of the actual computational needs or workload variations over time. While static allocation ensures resource isolation and predictability, it falls short in optimizing resource usage and adapting to the fluctuating demands typical of research activities. This inefficiency is particularly evident when GPUs remain underutilized despite the presence of user demands that needs additional resources. To address these challenges, we need more flexible and dynamic GPU allocation mechanisms that allow GPUs to be allocated or re-allocated to users' compute instances based on real-time demand and workload characteristics. 
Once a dynamic mechanism for GPU allocation is realized, it aligns well with the characteristics of the adaptable calendar-based system discussed above, further enhancing the ability to match resource availability with users' varying needs.

Finally, some NVIDIA GPUs support vGPU technology, which enables multiple Virtual Machines (VMs) to share a GPU transparently, making it suitable for research environments where full GPU capacity is not always required. However, vGPU licensing is not free and includes different type of virtualization products (e.g., vApps, vPC, vWS, vComputeServer) that specify user limits and features. This can introduce additional costs and planning problem, particularly when scaling GPU access across multiple users or nodes in research settings.  

\subsection{Sharing programmable switches in federated testbeds}

Programmable switches unveiled the ability of the network to offload generic computation to network switches and accelerate numerous tasks, a paradigm known as ``in-network computing''. In-network computing has shown dramatic improvements both for infrastructure-related services, such as telemetry, as well as for several distributed applications, such as key-values stores, consensus protocols, congestion control, distributed deep learning, among others.  


P4~\cite{p4} is a state-of-the-art abstraction to program high performance switches. In the federated testbed that we  consider, described in Sec.~\ref{sec:restart}, P4-based switches are shared in clusters to enable reuse across the scientific community in Italy. 
Given the multi-tenancy, federated nature of the cluster, isolation mechanisms are essential to prevent tenants from interfering with each other and to ensure the ability to run concurrent experiments. 

Differently from general-purpose compute infrastructure, existing programmable data-plane technologies lack the essential support for multi-tenancy. This is also the case for the Intel Tofino~\cite{tofino} products line, supporting P4. These switches are equipped with multiple packet processing pipelines that can be programmed individually, thus in principle enabling some form of parallel program execution. However, they do not implement any mechanism for fault, resource and performance isolation across programs of different tenants, which can interfere with each other, either by accident or maliciously. For example, every time a P4 program is installed, all pipelines undergo a complete reconfiguration, potentially causing disruption to the operations of other tenants' programs.

The goal of this contribution is to provide a solution that allows tenants to deploy their own data-plane applications in a secure and isolated way.

\subsection{Existing gaps and our contributions}

Based on the above analysis, we summarize the main gaps that we address and then highlight our contributions as follows.

\subsubsection {Sharing resources in general}
Traditional resource management approaches based on calendars with time slots typically lead to under-utilization of resources and wasted capacity. A more \textit{adaptable calendar-based system} is needed to manage reservations. Unlike traditional time-slot-based methods, which tend to be rigid, this adaptable calendar system can dynamical interact with users simplifying the process of releasing unused booked resources, notifying the users of the last minute availability of resources and continuously monitoring that the allocated resources are being actually used. This system allow users to book resources in advance while still permitting the system to adjust allocations as conditions and needs change. 

\subsubsection {GPU sharing}
Current open-source virtualization solutions for GPU resource allocation rely on static partitioning methods. We need more flexible and GPU allocation mechanisms that allow GPUs to be allocated or re-allocated to users' compute instances (VMs and containers) dynamically on the fly, following the real-time users' demand.

\subsubsection{Sharing programmable networking hardware}

Programmable networking hardware lacks the essential support for multi-tenancy. We need mechanisms to support a parallel access to such resources where possible, or at least to coordinate sequential (time-slot based) access in a smooth way.

\subsubsection{Our contributions}

In this work, we first introduce \textit{SHARY (SHaring Any Resource made easY)}, a platform
designed to facilitate the resource reservation process in federated environments. SHARY offers a user-friendly and flexible interface for managing reservations of various resources, including GPUs, networking devices, storage, and general compute resources. It builds on the adaptable calendar-based system discussed earlier, integrating dynamic allocation mechanisms to support a wide range of experimental requirements. 
The approach combines the predictability of advance reservations with the flexibility required to adapt to varying resource demands, ensuring an optimal balance between planning and adaptability. We then present the \textit{Federated Infrastructure for GPU Orchestration (FIGO)}, aimed at improving GPU availability and utilization among researchers. FIGO addresses the limitations of static allocation methods by introducing a dynamic allocation model. This approach improves the efficiency of GPU usage, making high-performance computing resources more accessible to a wider range of researchers. Last but not least, we introduce the reservation management system of SUP4RNET  cluster of programmable switches, which permits governing the access of different tenants to a set of shared hardware switching resources available.

By integrating SHARY with FIGO and SUP4RNET, we strive to create a comprehensive and adaptive resource-sharing ecosystem, meeting the evolving needs of researchers and enabling more effective use of federated testbed infrastructures.



\section{Distributed testbed in RESTART}\label{sec:restart}

\begin{figure}[bt!]
    \centering
    \includegraphics[width=8cm]{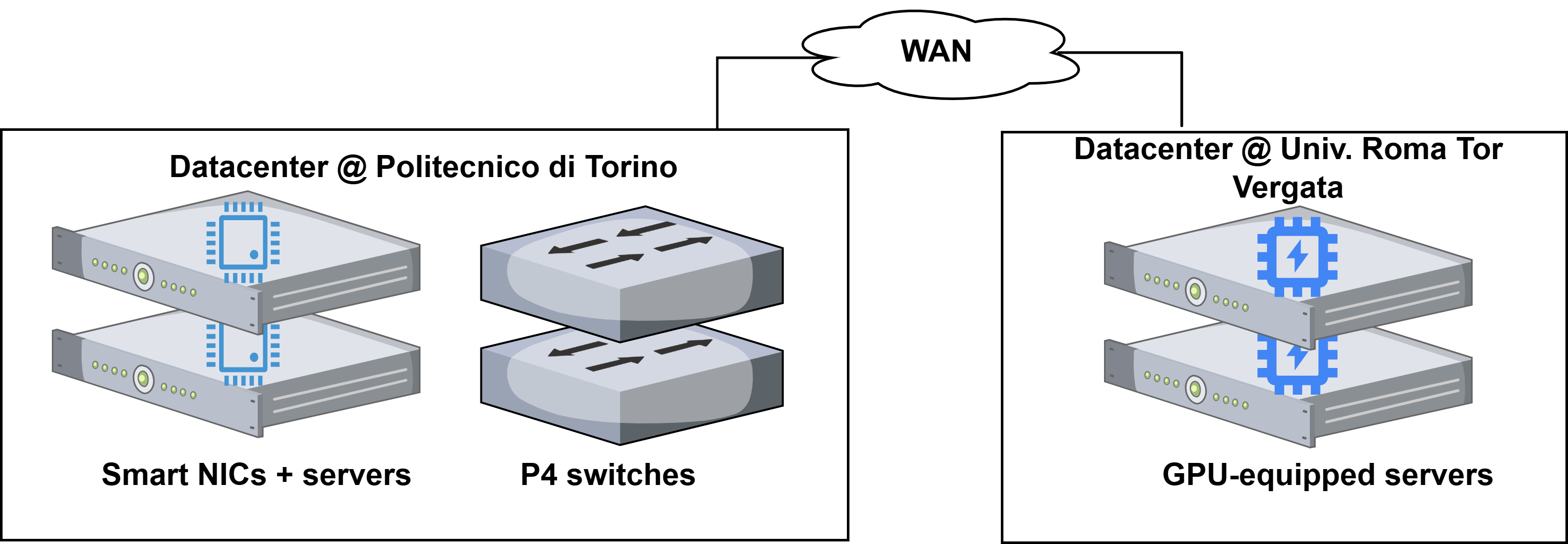}
    \caption{Federated testbed funded through the RESTART project} 
    \label{fig:testbed}
\end{figure}

A motivating scenario for our work is the federated testbed funded by RESTART project~\cite{restartWebsite}, whose deployment has just began at the time of writing. We describe here the parts of the resources that are currently included in the testbed, shown in Fig.~\ref{fig:testbed}.
We have currently two main sites, corrected through a WAN network. Additional sites are planned to be connected in the future. The first site is located in the main datacenter of Politecnico di Torino and the second site in datacenter of University of Rome Tor Vergata. 

In the site at the Politecnico di Torino, two kinds of resources are available: programmable switches and smart linecards.
The programmable switches are 2 Edgecore Wedge 100BF-32X, each equipped with a 2-pipelines Intel Tofino forwarding ASIC. The switches have 32 front-panel QSFP ports at 100Gbps link speed. 
The  smart linecards are hosted on high performing servers and they offer a variety of different programmable architectures. The following smart NICs are available: 4 NVIDIA Connect-X7, 4 NVIDIA BlueField 2, 2 NVIDIA Convergent accelerators with  GPU  A30X integrated with the BlueField 2, 4 AMD Alveo U45N data center accelerator cards, 2 AMD VCK5000 Versal development cards, 2 Intel Infrastructure Processing Unit F2000X-PL. The site of University of Rome Tor Vergata hosts a cluster of 4 GPUs NVIDIA L40S and 4 GPUs NVIDIA A16. 

Work is in progress to extend the RESTART testbed by federating further sites and bringing their resources to be shared.


\section{SHARY Sharing Platform}
\label{sec:shary}

\begin{figure}[b!]
    \centering
    \includegraphics[width=8cm]{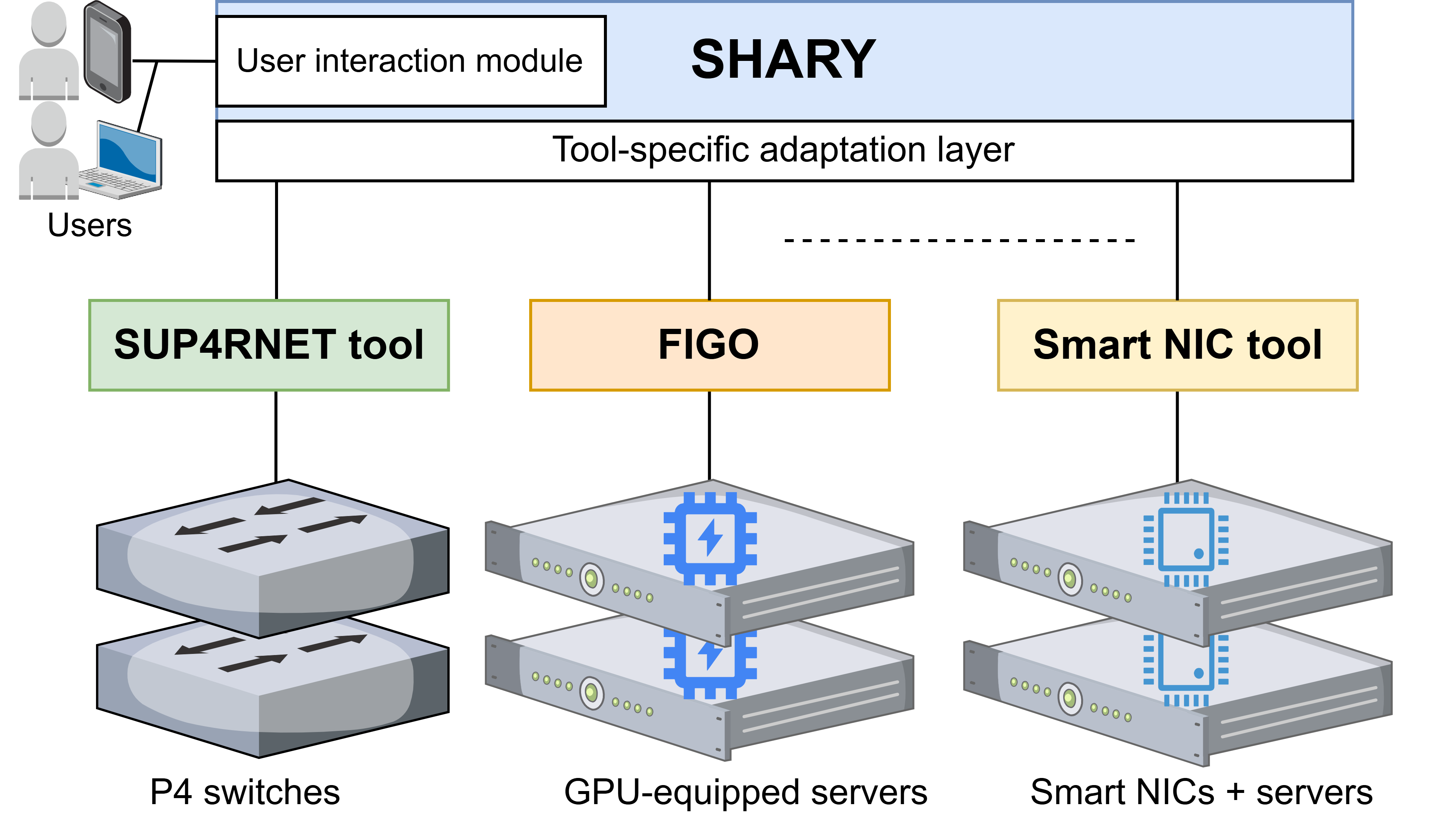}
    \caption{SHARY architecture to share heterogenous resources } 
    \label{fig:shary-architecture}
\end{figure}

\textit{SHARY (SHaring Any Resource made easY)} is a tool designed to facilitate the dynamic reservation of various types of resources in a federated environment. It aims to provide a user-friendly and adaptable system that simplifies the process of reserving, accessing, and managing resources such as GPUs, storage, networking devices, and compute nodes. SHARY enables researchers to interact with a web-based reservation system or programmatically via APIs, offering flexibility while maintaining control over resource access. The main features of SHARY are outlined in Section~\ref{sec:shary-reqs} and the main architecture is shown in Fig.~\ref{fig:shary-architecture}.

The SHARY home page is hosted on GitHub~\cite{github-shary}. It includes the updated documentation, the links to the repositories for the source code of the SHARY platform, and the links to the home pages of the related project (e.g., FIGO, SUP4RNET).  

\begin{figure*}[ht!]
    \centering
    \includegraphics[width=0.8\textwidth]{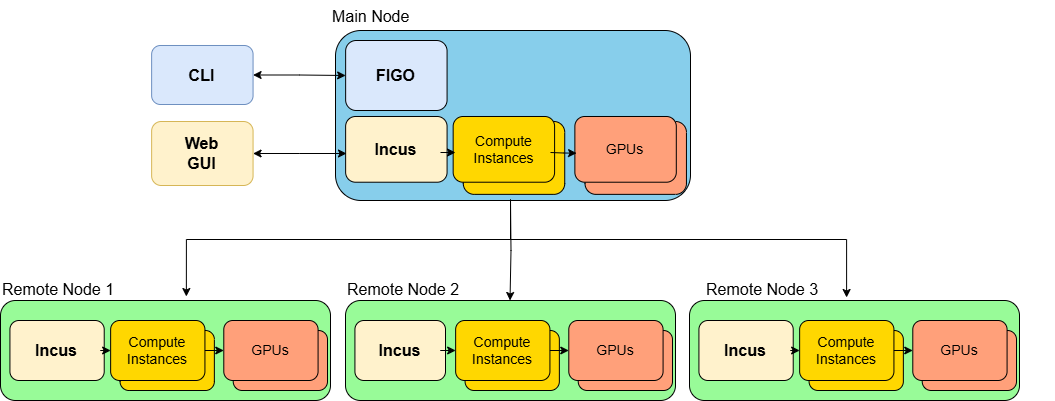}
    \caption{FIGO architecture - Each Remote Node runs an instance of incus, coordinated by the Main Node}
    \label{fig:figo-architecture}
\end{figure*}

\subsection{SHARY main features}
\label{sec:shary-reqs}

\begin{itemize}
    \item \textbf{Web GUI for User Reservation}: SHARY provides an easy-to-use web-based interface for resource reservation, with intuitive navigation and scheduling functionalities to facilitate user interaction.
    
    \item \textbf{User Interaction}: To allow a dynamic resource allocation and maximize utilization, SHARY  interacts with the users in order to incentivize the actual use of the allocated resources during the reservation periods, and the release of resources when not really used for the experimentations. SHARY can interact with the users through different communication means (email, social networks) to guarantee a prompt reaction from the users. 
    
    \item \textbf{API for Programmatic Access}: An API is offered to researchers who wish to interact with SHARY programmatically, allowing for automation and integration into custom workflows. This enables users to script interactions with the reservation system, adapt it to their specific needs, and integrate SHARY into larger automated research setups.

    \item \textbf{Standardized Batch Workload Procedures}: SHARY provides a standardized approach for managing batch workloads, ensuring consistency in submitting, queuing, and executing large-scale computational tasks. This simplifies the process for researchers to handle batch jobs, making it easier to manage their computational workloads efficiently.

    \item \textbf{Language to Express Reservation Policies}: A specific language should be provided to allow users and administrators to define reservation policies, including rules for access priority, availability, and preemptions. A web GUI front-end should complement this language, especially to enable end users to define and adjust these policies through an intuitive interface, making policy management accessible even for those without advanced technical skills.

    \item \textbf{Language to Describe Resources}: SHARY should support a standardized language that administrators can use to describe and enroll resources into the system. This language should be mapped into a user-friendly format to help users discover and understand the types of resources available for reservation. A web GUI front-end should complement this language, providing an intuitive interface that allows users to explore resource details visually. For advanced users, the same language should be accessible through the API, allowing them to specify detailed resource requirements when making reservations programmatically.

    \item \textbf{Differentiation Between Batch and Interactive Sessions}: SHARY can differentiate between batch and interactive sessions to better align with the needs of different research activities, enabling suitable scheduling and prioritization for each type of session.
        
    \item \textbf{Ownership Awareness}: SHARY can take into account the ownership of resources, ensuring that resource owners have priority and control over their assets, including the ability to reclaim resources when needed.
    
    \item \textbf{Prioritization with Reservation Advance Intervals}: The system enables prioritization of users by providing different reservation advance intervals, allowing users with higher priority or longer-term projects to book resources further in advance.
    
    \item \textbf{Notification of Last-Minute Availability}: SHARY can provide mechanisms to notify users if resources become available at the last minute, helping them seize opportunities for using newly freed-up resources.
    
    \item \textbf{Auction-like Mechanism}: The system can include an auction-like mechanism for resource allocation, allowing users to bid for access to limited resources based on their priority or urgency.
    
    \item \textbf{Tokens for Matching Actual Usage with Requested Usage}: To encourage users to request resources more accurately, a token system can be implemented where users gain tokens for aligning their actual usage with their reserved time, incentivizing efficient use of resources.

    \item \textbf{Smart Monitoring System}: SHARY integrates smart monitoring capabilities that can distinguish between development, batch workloads, and periods of inactivity across all types of resources. This monitoring system can provide valuable insights into current utilization patterns, helping to inform decisions that optimize resource allocation based on the nature of the task.
    
    \item \textbf{Comprehensive Accounting System}: SHARY integrates a detailed accounting system that tracks resource utilization, including GPU, CPU and programmable switch usage, disk space, and energy consumption. This ensures accurate monitoring, billing, and transparency in resource usage for all types of resources.
    
    \item \textbf{Visualization Front-end for Accounting}: A visualization tool is available for users and administrators to easily monitor resource usage and analyze data for optimization. This allows stakeholders to identify usage patterns and make data-driven decisions about resource allocation.

    \item \textbf{User Education and Training}: SHARY will include comprehensive user documentation and training modules to educate users on effectively utilizing the reservation system, the API, and the languages for resource description and policy definition. This will ensure that users at all skill levels can take full advantage of the system’s capabilities and reduce the learning curve for new users.
    
    \item \textbf{Modular control of heterogenous resources}: SHARY provides an adaptation layer to drive the specific management tools adopted to each kind of resource. E.g., in the Fig.~\ref{fig:shary-architecture} we show SHARY controlling the reservation on three kinds of resources: GPUs (throughout FIGO tool), P4 switches (throughout SUP4RNET tool) and smart NICs (the design of the related tool is ongoing).

    \item \textbf{Interoperability with External Tools}: (nice to have, for further study) it should be possible to integrate SHARY with other resource reservation systems and testbed platforms, such as FABRIC, CloudLab, and Fed4FIRE+. This interoperability would ensure that SHARY can be part of a larger ecosystem of federated research infrastructure, allowing users to access a broader range of resources and enabling collaborative experiments across different platforms.

\end{itemize}


\section{FIGO: Federated Infrastructure for GPU Orchestration}
\label{sec:figo}

\textit{FIGO (Federated Infrastructure for GPU Orchestration)} is a specialized system designed to optimize the sharing and utilization of GPUs within a federated research environment. Building upon the dynamic reservation capabilities provided by SHARY, FIGO focuses specifically on addressing the challenges of managing GPU resources across multiple sites. Its goal is to ensure that GPUs are allocated efficiently, enabling researchers to maximize the availability and performance of high-performance computing resources for AI and machine learning tasks. The requirements detailed in Section~\ref{sec:figo-reqs} describe the specific capabilities that FIGO must support to achieve effective GPU orchestration. FIGO is an open source project, the repository is on github~\cite{figo-doc}. The FIGO online documentation is available at~\cite{github-figo}.

\subsection{Main features of FIGO}
\label{sec:figo-reqs}

FIGO builds upon the dynamic reservation and resource management capabilities provided by SHARY (see Section~\ref{sec:shary-reqs}), focusing on the specialized needs of GPU orchestration within the federated environment. While SHARY provides the foundational features such as the comprehensive accounting system, batch workload procedures, and smart monitoring for various resources, FIGO extends these capabilities to address GPU-specific challenges. The main features of FIGO are:

\begin{itemize}
    \item \textbf{Pre-built Images of VMs and Containers}: FIGO offers access to pre-built VM and container images with AI tools pre-installed, simplifying the setup process for GPU-based research and reducing setup time for researchers.
    
    \item \textbf{Adaptation and Management of Heterogeneous GPU Resources}: FIGO manages the integration of different types of GPUs, varying numbers of GPU cores, memory capacities, and specific hardware features, ensuring that these resources are optimally allocated and utilized across the federated network.
    
    \item \textbf{Enhanced Smart Monitoring for GPUs}: Building on SHARY’s smart monitoring capabilities, FIGO includes GPU-specific monitoring features that track GPU performance metrics, utilization patterns, and temperature. This helps to identify bottlenecks and optimize the allocation of GPUs based on the specific requirements of AI and machine learning tasks.
    
    \item \textbf{GPU-Specific Batch Workload Optimization}: While SHARY provides standardized batch workload procedures, FIGO focuses on optimizing these procedures for GPU-intensive workloads, ensuring that large-scale GPU-based batch jobs are executed efficiently, with considerations for GPU parallelism and load balancing.
    
    \item \textbf{Support for Seamless Remote Interconnection}: FIGO facilitates seamless remote interconnection of GPUs with other resources, enabling distributed GPU workloads across different sites, and enhancing collaboration and computational capacity for AI research.
\end{itemize}


\subsection{FIGO Architecture and CLI commands}
\label{sec:figo-arch}
Figo is based on the Incus \cite{incusProject} manager of containers and virtual machine. 
Figo provides a coordination layer on top of Incus.
Figo is based on a straightforward architecture as it is running on a single main node, controlling a set of remote nodes through Incus; in fact, an instance of Incus is running both on the main node and on all the remote nodes. The architecture of FIGO is shown in Fig.~\ref{fig:figo-architecture}.


FIGO has been implemented as
a command-line tool designed to manage federated testbeds that combine CPU and GPU resources. It supports both virtual machines (VMs) and containers, making it suitable for distributed computing environments where centralized control is needed. FIGO simplifies tasks such as starting, stopping, and managing instances, integrating remote servers, and configuring user access through SSH and VPN setups. It is particularly useful for research and development scenarios requiring high-performance computing, like AI and machine learning, as it offers advanced GPU management and profile configuration features. FIGO’s comprehensive command suite helps reduce administrative overhead and ensures optimal use of computational resources.

The table below summarizes the key commands available in FIGO:

\begin{table}[h]
    \centering
    \caption{FIGO Command-Line Interface (CLI) Commands}
    \begin{tabular}{|l|p{6cm}|}
        \hline
        \textbf{Command} & \textbf{Description} \\
        \hline
        \texttt{figo instance} & Manage instances (VMs and containers), including starting, stopping, creating, and deleting. \\
        \hline
        \texttt{figo gpu} & Manage GPU profiles, such as adding or removing GPU resources to instances. \\
        \hline
        \texttt{figo profile} & Manage instance profiles, including listing, copying, and deleting profile configurations. \\
        \hline
        \texttt{figo user} & Handle user management, including adding new users, editing details, and managing access. \\
        \hline
        \texttt{figo remote} & Manage remote servers, including enrolling new nodes for centralized control. \\
        \hline
        \texttt{figo project} & Manage projects, such as creating and deleting isolated environments for different users or groups. \\
        \hline
        \texttt{figo vpn} & Configure VPN routes for secure communication between nodes in the federation. \\
        \hline
    \end{tabular}
    \label{tab:figo-commands}
\end{table}

Table \ref{tab:figo-commands} provides a summary of the available FIGO commands and their functions, illustrating the tool’s versatility in managing a federated compute infrastructure.


\section{Federating P4 switches with SUP4RNET}

\begin{figure}
    \centering
    \includegraphics[width=0.5\textwidth]{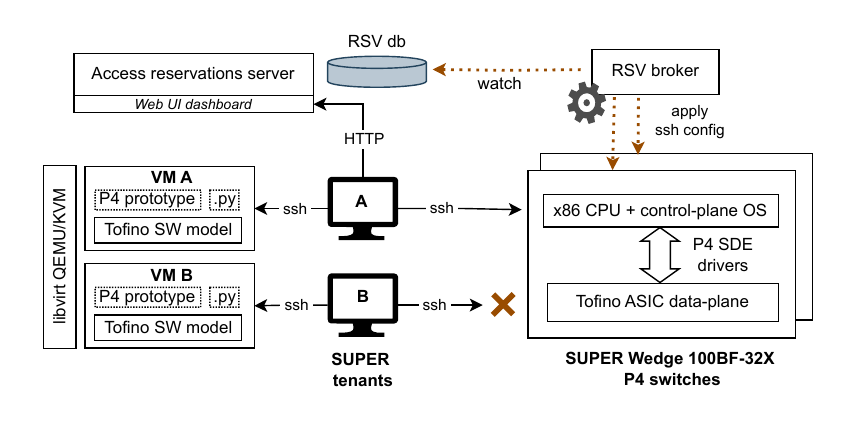}
    \caption{Architecture of the SUP4RNET P4 cluster and enabled P4 development workflow.}
    \label{fig:polito:p4cluster}
\end{figure}

We now describe the solution we adopted to permit governing the access of different tenants to the shared hardware switching resources available in the SUP4RNET cluster.

Fig.~\ref{fig:polito:p4cluster} shows the components we configured to allow tenants to experiment with P4, and the interactions among them in a typical development workflow.
The cluster readily supports the creation of per-tenant VM(s) pre-configured with Intel P4 Software Development Environment
 (SDE). We used \texttt{libvirt} to manage {QEMU/KVM}~\cite{qemu,kvm} virtual machines.

From a tenant's standpoint, compiling the P4 code and verifying its correctness for a Tofino target can be done entirely within a VM, even without accessing the physical switch. Indeed, the Intel P4 SDE provided within the base VM image provides a software behavioral model of the Tofino chip, implementing most of the Tofino functionalities and reproduces the pipeline behavior at register-level. 
When desired, tenants can run experiments on one or both the physical Tofino switches,
by explicitly reserving their access via a dedicated reservation web-site we developed. The Wedge 100BF-32X switches comes with a control-plane CPU where we installed a standard Linux-based distribution (i.e., Ubuntu), to serve as the switch OS. 
The switch OS manages the Tofino ASIC using the drivers provided with the Intel P4 SDE, which we installed in the switch CPU.
Every tenant is provided with a standard Linux user account on the switch OS, thus it can use this account to access the switch via \texttt{ssh}. 
 Permissions to login to the switch OS are dynamically granted and removed based on the reservations made by the tenants. A reservation (RSV) broker agent watches for changes to the RSV database and enforces a consistent configuration of the \texttt{ssh} permissions on the target switches. The RSV broker decouples the web server logic from the application of the permission configuration: manual changes to the RSV database entries are applied to the switches, independently of the web server component. Users \texttt{ssh} session and processes running after the end of the reserved time slot are considered as best effort, and can be terminated at any time by the RSV broker once a new valid reservation is made.
We used standard Linux groups and \texttt{sshd} configuration to implement the access control mechanism, and the utility \texttt{entr} to watch for changes to the RSV database.

We implemented an initial, but already operational solution to manage the access to Tofino switches in the SUP4RNET cluster. 
The SUP4RNET cluster is currently managed with Ansible, which fully automate VM creation, network configuration and user permissions. 

\section{Related Work}

\subsection{Federated Testbed Initiatives}

FABRIC (Adaptive ProgrammaBle Research Infrastructure for Computer Science and Science Applications) supports large-scale, advanced experiments in networking, cybersecurity, distributed computing, and other fields \cite{yankov2021fabric}. Launched in 2019, FABRIC builds on the GENI (Global Environment for Network Innovations) testbed, offering expanded capabilities \cite{baldin2020genifabric}. GENI, operational for over a decade, enabled extensive research in networking and distributed systems before ending in August 2023 \cite{fabric2024}. FABRIC's infrastructure includes 29 sites with substantial compute and storage resources, connected through high-speed optical links across commercial colocation facilities, national labs, and universities \cite{yankov2021fabric}. It integrates with specialized testbeds (e.g., 5G/IoT PAWR) and HPC centers, offering a rich environment for diverse research \cite{maccartney2021pawr}.

Fed4FIRE+ is an EU-funded federation of testbeds, active since 2017, that provides access to various networking testbeds across Europe \cite{de_turck2019fed4fire, serrano2017experiment}. It supports research in wireless networks, cloud computing, and IoT, offering tools for orchestration and monitoring, which help automate experiments and gather data. Its federated approach allows researchers to use resources across multiple sites, simplifying large-scale experimentation \cite{fed4fireWebsite}.

Emulab, active since 2000, offers controlled environments for network research \cite{white2002anatomy}. In 2014, it was extended through CloudLab, a virtualized platform for cloud computing experiments \cite{ricci2014introducing}. Both platforms integrate with GENI, providing broader experimental capabilities \cite{geniWebsite}.

\subsection{Specialized Platforms for 5G Research}

Colosseum is a large-scale wireless network emulator, specializing in 5G and Open RAN research \cite{colosseum2024tmc}. It allows high-fidelity simulation of radio environments and supports collaborative research across other testbeds \cite{colosseumWebsite}. Colosseum's capabilities extend to emulating complex radio scenarios, providing a virtual environment that can mimic diverse conditions, such as urban or rural mobile networks.

\subsection{Experimental Platforms for Wireless and LTE Research}

CorteXlab \cite{massouri2014cortexlab} and the NITOS Future Internet Facility \cite{2014nitos} offer specialized environments for indoor Wi-Fi and LTE research. These platforms enable controlled testing of wireless protocols and network behaviors, providing detailed insights into signal propagation and interference patterns. CorteXlab focuses on software-defined radio experiments, while NITOS is equipped for evaluating next-generation wireless networks and IoT applications. Their emphasis on controlled, small-scale wireless experimentation makes them essential tools for advancing indoor network research.

\subsection{GPU Resource Management for Collaborative Research}

TensorHive \cite{tensorhive} is an open-source tool for managing and monitoring GPU resources in multi-user environments. It enables efficient scheduling and reservation of GPU resources, allowing researchers to manage compute tasks collaboratively. TensorHive supports functionalities such as job scheduling, automated notifications, and access management. It is particularly suitable for research institutions where effective sharing and monitoring of high-performance computing resources are critical.

\subsection{Multi-tenancy Support in Programmable Switches}

Several prior works, such as P4Visor~\cite{p4visor} and others~\cite{primeP4}, adopt compile-time merging to address the multi-tenancy challenge. With these approaches, the P4 programs of different tenants are statically combined and compiled to a single P4 binary, later installed on the target switch. The main challenge lies in the logic partitioning and the resource allocation for programs of different tenants, which may require active coordination cycles across tenants.

HyPer4~\cite{hyper4} is an example of solutions that use a hypervisor-like P4 program to dynamically emulate, i.e., being functionally equivalent to, other P4 programs. These approaches give the illusion of multiple data plane programs but incur significant overheads, limiting their applicability to software and FPGA targets.

Menshen~\cite{menshen} and P4VBox~\cite{p4vbox} propose extensions of the reconfigurable match-action (RMT) hardware architectures for data-plane multitenancy. Menshen adds an indirection layer, in the form of small tables, to lookup and load different per-tenant configurations dynamically for the same RMT resource, using a configuration identifier contained in the packets. These solutions allow running per-tenant packet processing logic; however, they are not designed for Tofino hardware.

A recent work is SwitchVM~\cite{switchVM}, which introduces a runtime interpreter to allow the execution of \emph{Data Plane Filters} (DPF) in a P4 sandbox environment. SwitchVM has been implemented on a Tofino target, making it a relevant approach for SUP4RNET. It enables time-sharing of the pipeline hardware across multiple tenants while ensuring strict resource isolation between DPFs. However, users must write in a DPF-custom syntax rather than P4.




\section{Conclusion}

Federated testbeds play a crucial role in advancing research across cloud computing, networking, AI, and related fields by providing access to diverse resources across multiple sites. However, efficiently managing these resources presents challenges, particularly with high-demand assets like GPUs and specialized networking devices such as programmable switches. This work addresses these challenges within the RESTART research program in Italy, aiming to improve resource-sharing adaptability and efficiency.

We highlighted key issues, including underutilization of specialized hardware, the need for dynamic resource allocation, and the complexity of coordinating diverse resources. Traditional methods like static partitioning for GPUs and time-based access for switches often lead to inefficient resource use. To overcome these, we introduced \textit{SHARY}, a platform for flexible resource reservations, integrating dynamic allocation with an adaptable calendar system.

\textit{FIGO} further enhances GPU orchestration, increasing availability and utilization across the research community by dynamically adapting to varying demands. In parallel, our approach to manage programmable switches in the SUP4RNET cluster enables better coordination, supporting both parallel and sequential access where needed.

By integrating SHARY, FIGO, and SUP4RNET management tool, this work creates a cohesive and adaptive resource-sharing ecosystem that aligns with the needs of researchers, optimizing the use of valuable infrastructure. This approach benefits Italy’s scientific community and serves as a model for improving resource management in federated environments worldwide.

\section{Acknowledgements}
This work was supported by the European Union - Next Generation EU under the Italian National Recovery and Resilience Plan (NRRP), Mission 4, Component 2, Investment 1.3, CUP E83C22004640001, partnership on “Telecommunications of the Future” (PE00000001 - program “RESTART”)

\ifCLASSOPTIONcaptionsoff
  \newpage
\fi



\bibliographystyle{IEEEtran}
\bibliography{references.bib}

\end{document}